\documentclass[aps,floatfix,preprint,showpacs,preprintnumbers,nofootinbib,superscriptaddress,natbib]{revtex4}

\usepackage{graphicx,float}
\usepackage{epsfig}		
\usepackage{dcolumn}
\usepackage{bm}
\usepackage{subfigure}

\usepackage[all]{xy}
\usepackage{amsmath,upgreek}
\usepackage{amssymb}

\usepackage{pdfpages}
\usepackage{color}
\usepackage{graphicx,epstopdf}
\usepackage[colorlinks=true,
 linkcolor=red,
 urlcolor=purple,
 citecolor=blue,hyperindex]{hyperref}
\def\0{\mbox{\tiny $0$}}
\def\1{\mbox{\tiny $1$}}
\def\2{\mbox{\tiny $2$}}
\def\3{\mbox{\tiny $3$}}
\def\4{\mbox{\tiny $4$}}
\def\5{\mbox{\tiny $5$}}
\def\6{\mbox{\tiny $6$}}
\def\7{\mbox{\tiny $7$}}
\def\8{\mbox{\tiny $8$}}
\def\9{\mbox{\tiny $9$}}

\def\f14{\mbox{\tiny $\frac{1}{4}$}}

\def\bb#1{\mbox{\footnotesize $(#1)$}}

\begin{document}

\title{Testing non-classicality with exact Wigner currents for an anharmonic quantum system}
\author{Alex E. Bernardini}
\email{alexeb@ufscar.br}
\affiliation{Departamento de F\'{\i}sica, Universidade Federal de S\~ao Carlos, PO Box 676, 13565-905, S\~ao Carlos, SP, Brasil.}
\date{\today}

\begin{abstract}
Phase-space features of the Wigner flow for an anharmonic quantum system driven by the harmonic oscillator potential modified by the addition of an inverse square (one-dimension Coulomb-like) contribution are analytically described in terms of Wigner functions and Wigner currents.
Reporting about three correlated continuity equations which quantify the flux of quantum information in the phase-space, the non-classicality profile of such an anharmonic system can be consistently obtained in terms of the fluxes of {\em probability}, {\em purity} and {\em von Neumann-like entropy}.
Considering that quantum fluctuations can be identified from distortions over the classical regime, they can be quantified through the above-mentioned information fluxes whenever some {\em classically bounded} volume of the phase-space is selected.
Our results suggest that the Wigner flow approach works as a probe of quantumness and classicality for a large set of anharmonic quantum systems driven by quantum wells.
\end{abstract}

\date{\today}
\maketitle

\section{Introduction}

The Weyl-Wigner \cite{Wigner} representation of quantum mechanics encompasses the phase-space dynamics of quantum systems so as to provide the straightforward access to several of their quantum information features without affecting the predictive power of quantum mechanics.
Even being much more appealing in the quantum scenario of optical physics \cite{Sch}, quite general aspects that circumvent the frontiers between classical and quantum descriptions of Nature can be more properly comprehended from such a phase-space formulation of quantum mechanics \cite{01A,02A,03A,Case,Steuernagel3}.

In this context, for the Weyl transform of a generic quantum operator, $\hat{O}$, defined by
\begin{equation}
O^W(q, p)
= \hspace{-.2cm} \int^{+\infty}_{-\infty} \hspace{-.15cm}ds\,\exp{\left[2\,i \,p\, s/\hbar\right]}\,\langle q - s | \hat{O} | q + s \rangle=\hspace{-.2cm} \int^{+\infty}_{-\infty} \hspace{-.15cm} dr \,\exp{\left[-2\, i \,q\, r/\hbar\right]}\,\langle p - r | \hat{O} | p + r\rangle,
\end{equation}
the Wigner function, $W(q, p)$, can be described as the Weyl transform of a density matrix operator, $\hat{\rho} = |\psi \rangle \langle \psi |$, as
\begin{equation}
 h^{-1} \hat{\rho} \to  W(q, p) =  (\pi\hbar)^{-1} 
\int^{+\infty}_{-\infty} \hspace{-.15cm}ds\,\exp{\left[2\, i \, p \,s/\hbar\right]}\,
\psi(q - s)\,\psi^{\ast}(q + s),
\end{equation}
which can also be read as the Fourier transform of the off-diagonal terms of the associated density matrix that, by the way, exhibits the properties of a real-valued {\em quasi}-probability distribution, since it can assume local negative values.
Of course, the phase-space formulation of quantum mechanics is not exclusively described by the Weyl-Wigner formalism. If, on one hand, it can be subsidized by the Moyal's picture of quantum mechanics \cite{Moyal}, where the noncommutative nature of coordinate and momentum operators supports the Moyal {\em star}-product definition through which the Weyl-Wigner formalism is recovered, on the other hand, Wigner functions cannot be strictly interpreted as probability distributions, so that alternative phase-space frameworks are admitted \cite{Ballentine, Husimi,Glauber,Sudarshan,Carmichael,Callaway} either to circumvent or even to elucidate the above-mentioned non-negative probability (mis)interpretation (cf. for instance the optical tomographic probability representation of quantum mechanics \cite{Amosov, Radon, Mancini} where the  Weyl-Wigner-Moyal equation is always positive, even for Wigner functions assuming negative values\footnote{
In the context of entropy and information dynamics, the associated symplectic tomographyc probability form of the Weyl-Wigner-Moyal equation works as a classical approach to quantum systems \cite{Mancini}.}).

Pragmatically, the Weyl transform and the Wigner function connect quantum observables, $\hat{O}$, with their respective expectation values by means of the trace of the product of the two operators, $\hat{\rho}$ and $\hat{O}$, evaluated according to the integral of the product of their Weyl transforms over all the phase-space volume \cite{Wigner,Case},
\begin{equation}
Tr_{\{q,p\}}\left[\hat{\rho}\hat{O}\right] \to \langle O \rangle = 
\int^{+\infty}_{-\infty} \hspace{-.15cm}\int^{+\infty}_{-\infty} \hspace{-.15cm} {dq\,dp}\,W(q, p)\,{O^W}(q, p),
\label{five}
\end{equation}
which is indeed consistent with a probability distribution interpretation supported by the normalization condition of $\hat{\rho}$, $Tr_{\{q,p\}}[\hat{\rho}]=1$.
Once that such statistical aspects related to the nature of the density matrix quantum operators are established, the Weyl-Wigner formalism also admits extensions from pure states to statistical mixtures where, for example, the purity $Tr_{\{q,p\}}[\hat{\rho}^2]$ is read as
\begin{equation}
Tr_{\{q,p\}}[\hat{\rho}^2] = 2\pi\int^{+\infty}_{-\infty} \hspace{-.15cm}\int^{+\infty}_{-\infty} \hspace{-.15cm} {dq\,dp}\,W(q, p)^2,
\label{pureza}
\end{equation}
with the $2\pi$ introduced so as to satisfy the constraints: $Tr_{\{q,p\}}[\hat{\rho}^2] = Tr_{\{q,p\}}[\hat{\rho}] = 1$, for pure states, which shall be relevant in the context of entropy and information dynamics.

More importantly, some Wigner related quantifiers of non-classicality measure, for instance, how far from each other are the quantum and classical descriptions of Nature. In particular, it has been demonstrated that such quantifiers can be constructed in terms of probability, entropy and purity fluxes, through their respective continuity equations expressed in terms of Wigner functions and Wigner currents \cite{EPL18,Entro02,Liouvillian}.
To test such quantifiers and verify their efficiency in describing quantum fluctuations from a departing classical regime, the anharmonic Hamiltonian quantum system driven by 
\begin{equation}\label{qua14}
H(q,\,p) = \frac{p^2}{2m} + \frac{m\,\omega^2}{2}q^2 + \frac{4\alpha^2-1}{8 \,m}\frac{\hbar^2}{q^2} - \alpha \hbar\omega,
\end{equation}
will be investigated along with the Weyl-Wigner framework, where the above introduced constant coefficients have been chosen in order to anticipate a simplifying dimensionless analysis of the problem.
The above Hamiltonian is particularly relevant in discussing the one-dimensional reduction of the Hydrogen atom Schr\"odinger equation, as well as for implementing typical scenarios of quantum cosmology \cite{JCAP18}.
From \eqref{qua14}, the evinced non-linear deviation from the harmonic oscillator profile, and its corresponding Wigner eigenfunctions, shall be discussed along this work in order to provide a singular tool kit for quantifying quantum from classical distortions and to test the general formalism for Wigner information fluxes \cite{EPL18}.

Our work is thus organized as follows.
In Section II, the fluid analog of the phase-space information fluxes associated to quantum entropy and purity quantifiers are recovered from the Weyl-Wigner formalism for quantum mechanics.
In particular, the recently discussed quantifiers for non-classicality (non-Liouvillian fluidity) \cite{EPL18} are recast in a dimensionless framework so as to be more workable for a larger prospect of Hamiltonian quantum systems.
In Section III, the classical profile for the anharmonic Hamiltonian system supported by Eq.~\eqref{qua14}, and the exact expressions for quantum fluctuations, given in terms of Wigner functions and Wigner currents, are all obtained.
In addition, a {\em bounce-like} model extension of the formalism is also considered.
The quantum distortions on the classical background, and the corresponding classical reduction, both quantified in terms of the Wigner information flux continuity equations, are discussed in Section IV. The results are shown to support the properties of the Wigner flow framework from Section II as an effective quantifier for non-classicality. 
Our conclusions are drawn in Section V and they emphasize the complementary aspects of the Wigner formalism in discussing boundaries between quantum and classical regimes.

\section{Phase-space flow analysis and continuity equations}

The dynamics of a (time dependent) Wigner function, $W(q,\,p;\,t)$, can be cast in the form of a vector flux, $\mathbf{J}(q,\,p;\,t)$, that describes the flow of $W(q,\,p;\,t)$ in the phase-space \cite{Steuernagel3,Ferraro11,Donoso12,Domcke}.
With the flow field, $\mathbf{J}(q,\,p;\,t)$, expressed by $\mathbf{J} = J_q\,\hat{q} + J_p\,\hat{p}$, where $\hat{p} = \hat{p}_q$, the quantum equivalent Liouville equation is given by the continuity equation \cite{Case,Ballentine,Steuernagel3,EPL18},
\begin{equation}
\frac{\partial W}{\partial t} + \frac{\partial J_q}{\partial q}+\frac{\partial J_p}{\partial p} \equiv
\frac{\partial W}{\partial t} + \mbox{\boldmath $\nabla$}\cdot \mathbf{J} =0,
\label{quaz51}
\end{equation}
with\begin{equation}
J_q(q,\,p;\,t)= \frac{p}{m}\,W(q,\,p;\,t), \label{quaz500BB}
\end{equation}
and
\begin{equation}
J_p(q,\,p;\,t) = -\sum_{\nu=0}^{\infty} \left(\frac{i\,\hbar}{2}\right)^{2\nu}\frac{1}{(2\nu+1)!} \, \left[\left(\frac{\partial~}{\partial q}\right)^{2\nu+1}\hspace{-.5cm}V(q)\right]\,\left(\frac{\partial~}{\partial p}\right)^{2\nu}\hspace{-.3cm}W(q,\,p;\,t),
\label{quaz500}
\end{equation}
where $V(q)$ is the potential, and the contributions from $j \geq 1$ in the series expansion depict the distortion due to the quantum features on the classical Liouvillian pattern.
For the generic discussion of non-relativistic quantum Hamiltonians like
\begin{equation}
H(q,\,p) = \frac{p^2}{2m} + V(q),
\end{equation}
where $m$ is the particle's mass, tremendously simplified results can be obtained when $H(q,\,p)$ is put into a dimensionless form, $\mathcal{H}(x,\,k) = k^2/2 + \mathcal{U}(x)$, with the introduction of the dimensionless variables, $x = \left(m\,\omega\,\hbar^{-1}\right)^{1/2} q$ and $k = \left(m\,\omega\,\hbar\right)^{-1/2}p$, and the identification of $\mathcal{H} = (\hbar \omega)^{-1} H$ and $\mathcal{U}(x) = (\hbar \omega)^{-1} V\left(\left(m\,\omega\,\hbar^{-1}\right)^{-1/2}x\right)$, where $\omega^{-1}$ is a time scale.
In this case, the Wigner function and its Wigner current components can be mapped into dimensionless quantities given by
\begin{eqnarray}
\mathcal{W}(x, \, k;\,\tau) &\equiv& \left(m\omega\hbar\right)^{1/2}\, W(q,\,p;\,t),\\
\mathcal{J}_x(x, \, k;\,\tau) &\equiv& m \,\, J_q(q,\,p;\,t),\\
\mathcal{J}_k(x, \, k;\,\tau) &\equiv& \omega^{-1}\, J_p(q,\,p;\,t),
\end{eqnarray}
explicitly written as
\small\begin{eqnarray}\label{DimW}
\mathcal{W}(x, \, k;\,\tau) &=&  \pi^{-1} \int^{+\infty}_{-\infty} \hspace{-.15cm}dy\,\exp{\left[2\, i \, k \,y\right]}\,\varphi(x - y;\,\tau)\,\varphi^{\ast}(x + y;\,\tau),\quad \mbox{with $y = \left(m\,\omega\,\hbar^{-1}\right)^{1/2} s$},\,\,\,\,\\
\label{DimWA}\mathcal{J}_x(x, \, k;\,\tau) &=& k\,\mathcal{W}(x, \, k;\,\tau)
,\\
\label{DimWB}\mathcal{J}_k(x, \, k;\,\tau) &=& -\sum_{\nu=0}^{\infty} \left(\frac{i}{2}\right)^{2\nu}\frac{1}{(2\nu+1)!} \, \left[\left(\frac{\partial~}{\partial x}\right)^{2\nu+1}\hspace{-.5cm}\mathcal{U}(x)\right]\,\left(\frac{\partial~}{\partial k}\right)^{2\nu}\hspace{-.3cm}\mathcal{W}(x, \, k;\,\tau),
\end{eqnarray}\normalsize
where $\tau = \omega t$ is the dimensionless time, $\varphi(x,\,\tau)$ is consistent with the normalization condition given by
\begin{equation}
\int^{+\infty}_{-\infty} \hspace{-.2 cm}{dx}\,\vert\varphi(x;\,\tau)\vert^2 =\int^{+\infty}_{-\infty}\hspace{-.2 cm}{dq}\,\vert\psi(q;\,t)\vert^2 = 1,
\end{equation}
and the Eq.~\eqref{quaz51} can be multiplied by $(m\hbar/\omega)^{1/2}$ so as to return the dimensionless continuity equation,
\begin{equation}
\frac{\partial \mathcal{W}}{\partial \tau} + \frac{\partial \mathcal{J}_x}{\partial x}+\frac{\partial \mathcal{J}_k}{\partial k} = \frac{\partial \mathcal{W}}{\partial \tau} + \mbox{\boldmath $\nabla$}_{\xi}\cdot\mbox{\boldmath $\mathcal{J}$} =0,
\end{equation}
where the phase-space coordinate vector, $\mbox{\boldmath $\xi$} = (x,\,k)$, is identified.
Once the framework has been established, the phase-space information flux extensions of the above continuity equation can then be obtained through some elementary mathematical manipulations \cite{EPL18}.

Departing from the subliminar properties of locally and globally conservative systems -- which are respectively associated to a point in the phase-space, $\mbox{\boldmath $\xi$}$, and to a phase-space {\em volume integral bounded by a comoving closed surface}, $V = \int_{V}dx\,dk$ -- a substancial derivative \cite{Steuernagel3,Gradshteyn} operator can be defined by
\begin{equation}
\frac{D~}{D\tau} \equiv \frac{\partial~}{\partial\tau} + \mathbf{v}_{\xi}\cdot\mbox{\boldmath $\nabla$}_{\xi}\label{quaz57C},
\end{equation}
with $\mathbf{v}_{\xi} = d{\mbox{\boldmath $\xi$}}/d\tau  = (v_x,\,v_k)$ corresponding to the phase-space velocity (not necessarely the classical one) along a two-dimensional path which encloses an element of volume, $V$.
Through Eq.~\eqref{quaz57C}, an equivalent version of the {\em rate of change theorem} (cf. Eq.~(10.811) in Ref.~\cite{Gradshteyn}) can be established
for the Wigner function as
\begin{equation}
\frac{D~}{D\tau} \int_{V}dV\,\mathcal{W} \equiv 
\int_{V}dV\,\left[\frac{D\mathcal{W}}{D\tau} +  \mathcal{W} \mbox{\boldmath $\nabla$}_{\xi}\cdot \mathbf{v}_{\xi}\right]\label{quaz57D},
\end{equation}
with $dV \equiv dx\,dk$. 
If $ \mathbf{v}_{\xi}$ is identified with the classical phase-space vector velocity, $\mathbf{v}_{\xi(\mathcal{C})} = (k,\, -\partial \mathcal{U}/\partial x)$, through which a two-dimensional classical path, $\mathcal{C}$, can be delineated, one has from Eq.~\eqref{quaz57C},
\begin{equation}
\frac{D \mathcal{W}}{D\tau} = - \mathcal{W}\, \mbox{\boldmath $\nabla$}_{\xi} \cdot \mathbf{v}_{\xi(\mathcal{C})},
\label{quaz57B}
\end{equation}
which implies into a conservation law, ${D \mathcal{W}}/{D\tau} = 0$, given that the divergenceless (Liouvillian) behavior of the classical fluid-analog of the flow of the Wigner function is identified by $\mbox{\boldmath $\nabla$}_{\xi} \cdot \mathbf{v}_{\xi(\mathcal{C})} = 0$.
Otherwise, for the quantum scenario, one has an {\em ansatz} for $\mbox{\boldmath$\mathcal{J}$}$, $\mbox{\boldmath$\mathcal{J}$} = \mathbf{w}\,\mathcal{W}$, with the Wigner phase-velocity, $\mathbf{w}$, satisfying the constraint given by
\begin{equation}
\mbox{\boldmath $\nabla$}_{\xi} \cdot \mathbf{w} = \frac{\mathcal{W}\, \mbox{\boldmath $\nabla$}_{\xi}\cdot \mbox{\boldmath$\mathcal{J}$} - \mbox{\boldmath$\mathcal{J}$}\cdot\mbox{\boldmath $\nabla$}_{\xi}\mathcal{W}}{\mathcal{W}^2} \neq0,
\label{quaz59}
\end{equation}
which is translated into a non-Liouvillian behavior \cite{Steuernagel3,EPL18}, and for which it has been noticed that $\mbox{\boldmath $\nabla$}_{\xi}\cdot\mbox{\boldmath$\mathcal{J}$} = \mathcal{W}\,\mbox{\boldmath $\nabla$}_{\xi}\cdot\mathbf{w}+ \mathbf{w}\cdot \mbox{\boldmath $\nabla$}_{\xi}\mathcal{W}$.

The above elements, as established in \cite{EPL18}, can be helpful in discussing the quantum nature of Hamiltonians that describe periodic motions (driven by some kind of potential well). The periodic motion, in this case, is mapped into a phase-space two-dimensional volume enclosed by a classical path, $\mathcal{C}$, for which the phase-space volume integrated probability flux enclosed by $\mathcal{C}$, $\sigma_{(\mathcal{C})}$, can be identified by
\begin{equation}
\sigma_{(\mathcal{C})} =\int_{V_{_{\mathcal{C}}}}dV\,\mathcal{W}.
\label{quaz60}
\end{equation}
From a straightforward manipulation involving Eqs.~(\ref{quaz57C})-(\ref{quaz57B}), using the properties of $\mathbf{w}$, one also obtains
\begin{equation}
 \frac{D~}{D\tau}\sigma_{(\mathcal{C})} =\frac{D~}{D\tau}\int_{V_{_{\mathcal{C}}}}dV \,\mathcal{W} =  \int_{V_{_{\mathcal{C}}}}dV \,\left[\mbox{\boldmath $\nabla$}_{\xi}\cdot (\mathbf{v}_{\xi(\mathcal{C})}\mathcal{W}) - \mbox{\boldmath $\nabla$}_{\xi}\cdot \mbox{\boldmath$\mathcal{J}$}\right],
\label{quaz51CC}
\end{equation}
which allows for identifying the role of the quantum corrections
given in terms of $\Delta \mbox{\boldmath$\mathcal{J}$} = \mbox{\boldmath$\mathcal{J}$} - \mathbf{v}_{\xi(\mathcal{C})}\mathcal{W}$, which effectively drives the rate of change of $\sigma_{(\mathcal{C})}$ and the outward flux of $\mbox{\boldmath$\mathcal{J}$}$ (through $\mathcal{C}$), both in terms of a path integral given by
\begin{equation}
 \frac{D~}{D\tau}\sigma_{(\mathcal{C})} = -\int_{V_{_{\mathcal{C}}}}dV\,  \mbox{\boldmath $\nabla$}_{\xi}\cdot \Delta \mbox{\boldmath$\mathcal{J}$} = -\oint_{\mathcal{C}}d\ell\, \Delta\mbox{\boldmath$\mathcal{J}$}\cdot \mathbf{n}\equiv -\oint_{\mathcal{C}}d\ell\, \mbox{\boldmath$\mathcal{J}$}\cdot \mathbf{n},
\label{quaz51DD}
\end{equation}
where the unitary vector $\mathbf{n}$ is defined by $\mathbf{n}= (-d{k}_{_{\mathcal{C}}}/d\tau, d{x}_{_{\mathcal{C}}}/d\tau) \vert\mathbf{v}_{\xi(\mathcal{C})}\vert^{-1}$, such that, in the last step, one has noticed that $\mathbf{n}\cdot\mathbf{v}_{\xi(\mathcal{C})}= 0$. Therefore,
for the line element, $\ell$, set as $d\ell \equiv \vert\mathbf{v}_{\xi(\mathcal{C})}\vert d\tau$, one has a parametric integral given by
\begin{equation}
\frac{D~}{D\tau}\sigma_{(\mathcal{C})}
\bigg{\vert}_{\tau = T} = -\oint_{\mathcal{C}}d\ell\, \Delta\mbox{\boldmath$\mathcal{J}$}\cdot \mathbf{n} = -
\int_{0}^{T}d\tau\, \Delta \mathcal{J}_k(x_{_{\mathcal{C}}}\bb{\tau},\,k_{_{\mathcal{C}}}\bb{\tau};\tau)\,\,\frac{d}{d\tau}{x}_{_{\mathcal{C}}}\bb{\tau},
\label{quaz51EE}
\end{equation}
where $x_{_{\mathcal{C}}}\bb{\tau}$ and $k_{_{\mathcal{C}}}\bb{\tau}$ are typical solutions of the classical Hamiltonian problem, $T=2\pi$ is the dimensionless period of the classical motion, and $\Delta \mathcal{J}_k(x,\,p;\tau)$ is identified by the piece of the series expansion from Eq.~\eqref{quaz500} with $\nu\geq 1$.
Of course, the integral from Eq.~\eqref{quaz51DD} vanishes in the classical limit, i.e. for $\mbox{\boldmath$\mathcal{J}$} \sim \mathbf{v}_{\xi(\mathcal{C})}\mathcal{W}$. Therefore, Eq.~\eqref{quaz51EE} works as an optimized quantifier of non-classicality for a {\em pletora} of Wigner functions.

In order to extend the range of applicability of the above result, one identifies the Wigner related von Neumann entropy and  quantum purity respectively by \cite{EPL18}
\begin{equation}
{S}_{vN} =-\int_{V}dV\,\mathcal{W}\,\ln\vert \mathcal{W}\vert,
\label{quaz60}
\end{equation}
and
\begin{eqnarray}
\mathcal{P} = 2\pi \int_{V}dV\,\, \mathcal{W}^2,
\label{quaz63}
\end{eqnarray}
such that their temporal rate of change are respectively given by
\begin{eqnarray}
\frac{D{S}_{vN}}{D\tau} &=& -\frac{D~}{D\tau}\left(\int_{V}dV\, \mathcal{W}\,\ln(\mathcal{W})\right)
\nonumber\\
&=& -\int_{V}dV\,\left[\frac{D~}{D\tau} (\mathcal{W}\,\ln(\mathcal{W})) + \mathcal{W}\,\ln(\mathcal{W}) \mbox{\boldmath $\nabla$}_{\xi}\cdot \mathbf{v}_{\xi(\mathcal{C})}\right]\nonumber\\
&=&- \int_{V}dV\,\left[\frac{\partial~}{\partial \tau} (\mathcal{W}\,\ln(\mathcal{W})) + \mbox{\boldmath $\nabla$}_{\xi}\cdot(\mathbf{v}_{\xi(\mathcal{C})} \mathcal{W}\,\ln(\mathcal{W}))\right],
\label{quaz61}
\end{eqnarray}
and
\begin{eqnarray}
\frac{1}{2\pi}\frac{D\mathcal{P}}{D\tau} &=& \frac{D~}{D\tau}\left(\int_{V}dV\, \mathcal{W}^2\right)\nonumber\\
&=&
\int_{V}dV\,\left[\frac{D~}{D\tau} \mathcal{W}^2 + \mathcal{W}^2 \mbox{\boldmath $\nabla$}_{\xi}\cdot \mathbf{v}_{\xi(\mathcal{C})}\right]\nonumber\\
&=& \int_{V}dV\,\left[\frac{\partial~}{\partial \tau}\mathcal{W}^2 + \mbox{\boldmath $\nabla$}_{\xi}\cdot(\mathbf{v}_{\xi(\mathcal{C})} \mathcal{W}^2)\right]
\label{quaz64}
\end{eqnarray}
from which, after noticing that $\partial \mathcal{W}/\partial \tau = - \mbox{\boldmath $\nabla$}_{\xi}\cdot\mbox{\boldmath$\mathcal{J}$} = - \mbox{\boldmath $\nabla$}_{\xi}\cdot (\mathbf{w}\,\mathcal{W}$), and using Eq.~\eqref{quaz59}, one obtains, respectively,
\begin{eqnarray}
\frac{D{S}_{vN}}{D\tau} &=& \int_{V}dV\,\left[\mathcal{W} \,\mbox{\boldmath $\nabla$}_{\xi}\cdot\mathbf{w} +
 \mbox{\boldmath $\nabla$}_{\xi}\cdot\left(\mbox{\boldmath$\mathcal{J}$} \ln(\mathcal{W}) - \mathbf{v}_{\xi(\mathcal{C})} \mathcal{W}\,\ln(\mathcal{W})\right)\right]\nonumber\\
 &=& \int_{V}dV\,\mathcal{W} \,\mbox{\boldmath $\nabla$}_{\xi}\cdot\mathbf{w}
+ \oint_{\mathcal{\,\,}}d\ell\, \ln(\mathcal{W})\left(\mbox{\boldmath$\Delta\mathcal{J}$}\cdot \mathbf{n}\right)\nonumber\\
^{V\to\infty} &=& \langle \mbox{\boldmath $\nabla$}_{\xi}\cdot\mathbf{w}\rangle,
\label{quaz62}
\end{eqnarray}
and
\begin{eqnarray}
\frac{1}{2\pi}\frac{D\mathcal{P}}{D\tau} &=& -\int_{V}dV\,\left[\mathcal{W}^2 \,\mbox{\boldmath $\nabla$}_{\xi}\cdot\mathbf{w} +
 \mbox{\boldmath $\nabla$}_{\xi}\cdot(\mbox{\boldmath$\mathcal{J}$} \mathcal{W} - \mathbf{v}_{\xi(\mathcal{C})} \mathcal{W}^2)\right]\nonumber\\
&=& \int_{V}dV\,\mathcal{W}^2 \,\mbox{\boldmath $\nabla$}_{\xi}\cdot\mathbf{w}
+ \oint_{\mathcal{\,\,}}d\ell\, \mathcal{W} \left(\mbox{\boldmath$\Delta\mathcal{J}$}\cdot \mathbf{n}\right)\nonumber\\
^{V\to\infty} &=& \langle \mathcal{W} \mbox{\boldmath $\nabla$}_{\xi}\cdot\mathbf{w}\rangle,
\label{quaz62BBB}
\end{eqnarray}
where $\langle \dots \rangle = Tr_{\{x,k\}}\left[\hat{\rho}(\dots)\right]$, and the surface terms have been suppressed in the limit where $V \to \infty$, as to recover the results from \cite{EPL18}\footnote{After suitable mathematical manipulations involving the definitions from Eqs.~\eqref{DimW}-\eqref{DimWB}, it is possible to verify that $\mbox{\boldmath $\nabla$}_{\xi}\cdot\mathbf{w}$ is proportional to $$ \sum_{\nu=1}^{\infty} \left(\frac{i}{2}\right)^{2\nu}\frac{1}{(2\nu+1)!} \, \left[\left(\frac{\partial~}{\partial x}\right)^{2\nu+1}\hspace{-.4cm}\mathcal{U}\right]\,\frac{\partial~}{\partial k}\left((1/\mathcal{W})\frac{\partial~}{\partial k}\right)^{2\nu}\hspace{-.3cm}\mathcal{W},$$ from which one notices that symmetric potentials and parity-defined Wigner distributions both lead to vanishing values for the above obtained time derivatives.}.

For a finite volume, $V_{_{\mathcal{C}}}$, identified by that one enclosed by the classical surface, ${\mathcal{C}}$, the surface term must be reconsidered into the above calculation. Given that, for parity symmetric potentials, $\mathcal{U}(x) =\mathcal{U} (-x)$, driving periodic ((an)harmonic) oscillations, the above obtained averaged contributions vanish, the corresponding continuity equations can be respectively recast in the form of
\begin{eqnarray}
\frac{D~}{D\tau}{S}_{vN(\mathcal{C})}
\bigg{\vert}_{\tau = T} &=& \oint_{\mathcal{C}}d\ell\, \ln(\mathcal{W})\left(\Delta\mbox{\boldmath$\mathcal{J}$}\cdot \mathbf{n}\right) \nonumber\\
&=&
\int_{0}^{T}d\tau\, \ln(\mathcal{W}(x_{_{\mathcal{C}}}\bb{\tau},\,k_{_{\mathcal{C}}}\bb{\tau};\tau))\,\,\Delta \mathcal{J}_k(x_{_{\mathcal{C}}}\bb{\tau},\,k_{_{\mathcal{C}}}\bb{\tau};\tau)\,\,\frac{d~}{d\tau}{x}_{_{\mathcal{C}}}\bb{\tau},
\label{quaz62CC}
\end{eqnarray}
and
\begin{eqnarray}
\frac{D~}{D\tau}\mathcal{P}_{(\mathcal{C})}
\bigg{\vert}_{\tau = T} &=& -\oint_{\mathcal{C}}d\ell\, \mathcal{W}\,\Delta\mbox{\boldmath$\mathcal{J}$}\cdot \mathbf{n} \nonumber\\&=& -
\int_{0}^{T}d\tau\, \mathcal{W}(x_{_{\mathcal{C}}}\bb{\tau},\,k_{_{\mathcal{C}}}\bb{\tau};\tau) \,\Delta \mathcal{J}_k(x_{_{\mathcal{C}}}\bb{\tau},\,k_{_{\mathcal{C}}}\bb{\tau};\tau)\,\,\frac{d~}{d\tau}{x}_{_{\mathcal{C}}}\bb{\tau},
\label{quaz64DD}
\end{eqnarray}
with $\Delta\mbox{\boldmath$\mathcal{J}$}\cdot \mathbf{n} \equiv \mbox{\boldmath$\mathcal{J}$}\cdot \mathbf{n}$, through which one can quantify the quantum fluctuations that distinguish quantum from classical regimes whenever some classical boundary trajectory in the phase-space is specified. For quantum systems which account for the all order corrections from Eq.~\eqref{DimWB}, quantumness and classicality can thus be quantified through the above obtained continuity equation framework in terms of the results from Eqs.~\eqref{quaz51EE}, \eqref{quaz62CC}, and \eqref{quaz64DD}.

\section{Wigner function and Wigner currents for the $1$-dim harmonic oscillator plus inverse square potential}

For the quantum system from Eq.~\eqref{qua14}, the dimensionless Schr\"odinger equation is written as 
\begin{equation}\label{qua16}
\mathcal{H} \varphi^{\alpha}_n(x) = \frac{1}{2}\left\{-\frac{d^{2}}{dx^{2}}+ x^{2} + \frac{4 \alpha^2 -1}{4 x^{2}}-2\alpha \right\} \varphi^{\alpha}_n(x) = \varepsilon_n\,\varphi^{\alpha}_n({x}),
\end{equation}
where $k\equiv -i\,(d/dx)$, $\alpha$ is a continuous value parameter, and one identifies the quantum number $n$ as related to the self-energy, $E_n =\hbar \omega\, \varepsilon_n$, through $\varepsilon_n = 2n + 1$ (cf. Eq.~\eqref{qua14}).
The exact solution for the above quantum mechanical problem is given by
\begin{equation}
\varphi^{\alpha}_n(x) = 2^{{1}/{2}}\,\Theta(x) \, N_n^{(\alpha)}\, x^{\alpha + \frac{1}{2}}\,\exp(-x^2/2)\,L^{\alpha}_n(x^2),
\label{qua19}
\end{equation}
where $\Theta(x)$ is the {\em step-unity function} that constrains the result to $0 < x < \infty$, $L^{\alpha}_n$ are the {\em associated Laguerre polynomials}, and $N_n^{(\alpha)}$ is the normalization constant given by
\begin{equation}
N_n^{(\alpha)} = \sqrt{\frac{n!}{\Gamma(n+\alpha+1)}}, 
\label{qua20}
\end{equation}
where $\Gamma(n) = (n-1)!$ is the {\em gamma function}.
An approximated bounce model \cite{JCAP18} corresponding to an even symmetrization of the above solution, which should be valid for $-\infty < x < +\infty$, can be obtained by simply suppressing the {\em step-unity function}, $\Theta(x)$.

By substituting the stationary states, $\varphi^{\alpha}_n(x)$, into the dimensionless form of the Wigner function from Eq.~\eqref{DimW}, one obtains
\small\begin{eqnarray}\label{DimW2}
\mathcal{W}_n^{\alpha}(x, \, k) 
&=&2 (N_n^{(\alpha)})^2\,  \pi^{-1} \int^{+\infty}_{-\infty} \hspace{-.15cm}dy\,\Theta(x+y)\Theta(x-y)\,(x^2-y^2)^{\frac{1}{2}+\alpha}\, \exp\left(2\,i\, k\,y\right)\\
&&\qquad\qquad\qquad\qquad\qquad\qquad\qquad\exp\left[-(x^2+y^2)\right] L_n^{\alpha}\left((x+y)^2\right)\,L_n^{\alpha}\left((x-y)^2\right)
\nonumber\\
&=& \frac{2}{\pi} \int^{+x}_{-x} \hspace{-.15cm}dy\,\exp\left(2\,i\, k\,y\right)\,\exp\left[-(x^2+y^2)\right] \,\sum_{j=0}^n \frac{L_{n-j}^{\alpha+2j}\left(2 (x^2+y^2)\right) }{\Gamma(\alpha+j+1)}
\frac{(x^2-y^2)^{\frac{1}{2}+\alpha+2j}}{j!},\nonumber
\end{eqnarray}\normalsize
where it has been noticed that
\begin{equation}
L_n^{\alpha}\left(x\right)\,L_n^{\alpha}\left(y\right) = \frac{\Gamma(n+\alpha+1)}{n!}\,\sum_{j=0}^n \frac{L_{n-j}^{\alpha+2j}\left(x+y\right) }{\Gamma(\alpha+j+1)}
\frac{x^j y^j}{j!}.
\end{equation}
Given that the generating function of $L_n^{\alpha}\left(x\right)$ is given by
\begin{equation}
\frac{1}{(1-z)^{\alpha+1}}
\exp\left[-\frac{x\,z}{1-z}\right]= \sum_{n=0}^{\infty}L_n^{\alpha}\left(x\right)\,z^n,
\end{equation}
for $n\in$ integers, one can simply write
\small\begin{equation}
L_{n-j}^{\alpha+2j}\left(2 (x^2+y^2)\right) =\,\frac{1}{\Gamma(n-j+1)}\left(\frac{d~}{dz}\right)^{n-j}\left\{\frac{1}{(1-z)^{\alpha+1+2j}}
\exp\left[-\frac{2z(x^2+y^2)}{1-z}\right]\right\}\bigg{\vert}_{z=0},
\end{equation}\normalsize
which can then be substituted into Eq.~\eqref{DimW2} as to return
\footnotesize\begin{equation}\label{DimW3}
\mathcal{W}_n^{\alpha}(x, \, k) 
= \frac{4}{\pi} \sum_{j=0}^{n}
\left\{
\frac{x^{2(\alpha+1+2j)}}{(\alpha+j)!\,(n-j)!\,j!}
\left(\frac{d~}{dz}\right)^{n-j}
\left\{
\frac{1}{(1-z)^{\alpha+1+2j}}
\exp\left[-x^2\frac{1+z}{1-z}\right] \, \mathcal{G}^{(z)}_j(x)
\right\}\bigg{\vert}_{z=0}
\right\},
\end{equation}\normalsize
with
\begin{equation}
\mathcal{G}^{(z)}_j(x) = \int^{1}_{0} \hspace{-.15cm}dw\,\cos(2\, k\,x\,w)\,\exp\left[-x^2\frac{1+z}{1-z}w^2\right]
(1- w^2)^{\alpha+2j+1/2}.\nonumber
\end{equation}
Considering only the semi-integer values of $\alpha$ into the above expression, written as $\alpha = 1/2 + \upsilon$, with $\upsilon = 0,\,1,\,2,\, \dots$, one has the finite sum
\begin{equation}
(1-w^2)^{\alpha+2j+1/2} = (1-w^2)^{1 + \upsilon+2j} = \sum_{\ell = 0}^{1 + \upsilon+2j}
(-1)^\ell\,\frac{w^{2\ell}(1+\upsilon+2j)!}{(1+\upsilon+2j-\ell)!\,\ell!},
\end{equation}
which returns an expression for $\mathcal{G}^{(z)}_j(x)$ resumed by
\footnotesize\begin{eqnarray}
\label{finalform2}
\mathcal{G}^{(z)}_j(x) 
&=& \sum_{\ell = 0}^{\upsilon+1+2j}
\frac{x^{-(2\ell+1)}(\upsilon+1+2j)!}{(\upsilon+1+2j-\ell)!\,\ell!}
\left(\frac{d~}{d\mu}\right)^{\ell}\left\{\int^{1}_{0} \hspace{-.15cm}dw\,\cos(2\, k\,x\,w)\exp\left[-\mu\,x^2w^2\right]
\right\}\bigg{\vert}_{\mu=\frac{1+z}{1-z}}\\
&=&\frac{\sqrt{\pi}}{2} \sum_{\ell = 0}^{\upsilon+1+2j}
\frac{x^{-(2\ell+1)}(\upsilon+1+2j)!}{(\upsilon+1+2j-\ell)!\,\ell!}
\left(\frac{d~}{d\mu}\right)^{\ell}\left\{
\frac{1}{\sqrt{\mu}}
\exp\left(-\frac{k^{2}}{\mu}\right)\, \Re\left[\mbox{Erf}\left(\sqrt{\mu}x + i \frac{k}{\sqrt{\mu}}\right)\right]
\right\}\bigg{\vert}_{\mu=\frac{1+z}{1-z}}.\nonumber
\end{eqnarray}\normalsize
It provides a complete analytic expression for $\mathcal{W}_n^{\alpha}(x, \, k)$ given in terms of two finite series expansions.  

Now turning our attention to the computation of the dimensionless Wigner currents, the expression for the $x$-component is straightforwardly obtained from Eq.~\eqref{DimWA} as
\begin{equation}
\mathcal{J}_x^{n(\alpha)}(x,\,k) = k \,\mathcal{W}^{\alpha}_n(x,\,k).
\end{equation}
Correspondently, from Eq.~\eqref{DimWB}, the $k$-component,
\begin{eqnarray}
\mathcal{J}_k^{n(\alpha)}(x,\,k) = -\sum_{\nu=0}^{\infty} \left(\frac{i}{2}\right)^{2\nu}\frac{1}{(2\nu+1)!} \, \left[\left(\frac{\partial~}{\partial x}\right)^{2\nu+1}\hspace{-.5 cm}\mathcal{U}(x)\right]\,\left(\frac{\partial~}{\partial k}\right)^{2\nu} \hspace{-.3 cm}\mathcal{W}^{\alpha}_n(x,\,k),
\label{qua503}
\end{eqnarray}
provides quantum and non-linear corrections to the Liouvillian profile (cf. $\nu \geq 1$ contributions).
By following a careful manipulation partially reproduced from Ref.~\cite{JCAP18}, one first notices that the contribution from $\nu=0$, even if it includes non-linear terms, drives the classical (Liouvillian) behavior of the anharmonic system.
For the Hamiltonian system from Eq.~\eqref{qua14}, one thus should have 
\begin{eqnarray}
\mathcal{J}^{n(\alpha)}_{k(Cl)}(x,\,k)
&=& - \left(x + \frac{1-4\alpha^2}{4x^3}\right)\mathcal{W}^{\alpha}_n(x,\,k),
\label{qua505B}
\end{eqnarray}
for the classical limit. 
Naturally, the potential proportional to $x^2$ does not contribute to the quantum fluctuations since its contribution vanishes for $\nu \geq 1$.
Therefore, quantum fluctuations arise from the contribution due to the inverse square potential, $1/x^{2}$. In order to account for this contribution, one preliminarily notices that 
\begin{equation}
\left(\frac{\partial~}{\partial x}\right)^{2\nu+1}\frac{1}{x^2} = -(2\nu+2)\frac{(2\nu+1)!}{x^{2\nu+3}}
\label{qua68},
\end{equation}
and that
\begin{equation}
\left(\frac{\partial~}{\partial k}\right)^{2\nu}\mathcal{W}^{\alpha}_n(x,\,k) =
\frac{1}{\pi}\int_{-\infty}^{+\infty}dy\,(2\,i\,y)^{2\nu}\,\exp(2\,i\,k\,y)\,\varphi^{\alpha*}_n(x+y)\,\varphi_n^{\alpha}(x-y).
\label{qua69}
\end{equation}
One can then work out the sum in Eq.~(\ref{qua503}) for the term proportional to $1/x^{2}$ in $\mathcal{U}(x)$ as to obtain
\begin{eqnarray}
\lefteqn{-\sum_{\nu=0}^{\infty} 
\left(\frac{i}{2}\right)^{2\nu}\frac{1}{(2\nu+1)!} 
\left[\left(\frac{\partial~}{\partial x}\right)^{2\nu+1}
\frac{1}{x^2}\right]\,\left(\frac{\partial~}{\partial k}\right)^{2\nu}\, \mathcal{W}^{\alpha}_n(x,\,k)=}\nonumber\\
&&= \frac{2}{\pi x^3} 
\int_{-\infty}^{+\infty}dy\,\left[\sum_{\nu=0}^{\infty} (-1)^{2\nu}
\frac{(2\nu+1)!}{(2\nu+1)!}(\nu+1)
\left(\frac{y}{x}\right)^{2\nu}\right] \,\exp(2\,i\,k\,y)\,\varphi^{\alpha*}_n(x+y)\,\varphi_n^{\alpha}(x-y)
\nonumber\\
&&= \frac{2}{\pi x^3}\int_{-\infty}^{+\infty}dy\,\frac{d~}{d\kappa}
\left(\sum_{\nu=0}^{\infty} \kappa^{\nu+1}\right)
\,\exp(2\,i\,k\,y)\,\varphi^{\alpha*}_n(x+y)\,\varphi_n^{\alpha}(x-y)
\nonumber\\
&&= \frac{2 x}{\pi}\int_{-\infty}^{+\infty}dy\,(x^2-y^2)^{-2}\,\exp(2\,i\,k\,y) 
\,\varphi^{\alpha*}_n(x+y)\,\varphi_n^{\alpha}(x-y),
\label{qua504}
\end{eqnarray}
where $\kappa = y^2/x^2$, from which there is no restriction about considering $\kappa < 1$\footnote{Since the {\em step-unity function} contributions set $y \in (-x,+x)$.} such that, for the last step, it has been noticed that
$$\frac{d~}{d\kappa}\left(\sum_{k=0}^{\infty} \kappa^{k+1}\right) =\frac{d~}{d\kappa}\left(\sum_{k=1}^{\infty} \kappa^{k}\right) = (1-\kappa)^{-2}.$$
The final form of $\mathcal{J}^{n(\alpha)}_{k}(x,\,k)$ is thus given by
\begin{eqnarray}
\mathcal{J}^{n(\alpha)}_{k}(x,\,k)
&=& - x \left(\mathcal{W}^{\alpha}_n(x,\,k) + \frac{1-4\alpha^2}{4} \mathcal{Y}^{\alpha}_n(x,\,k)\right),
\label{qua505G}
\end{eqnarray}
with $\mathcal{W}^{\alpha}_n(x,\,k)$ given by Eq.~\eqref{DimW3}, and with
\footnotesize\begin{equation}\label{DimW6}
\mathcal{Y}_n^{\alpha}(x, \, k) 
= \frac{4}{\pi} \sum_{j=0}^{n}
\left\{
\frac{x^{2(\alpha-1+2j)}}{(\alpha+j)!\,(n-j)!\,j!}
\left(\frac{d~}{dz}\right)^{n-j}
\left\{
\frac{1}{(1-z)^{\alpha+1+2j}}
\exp\left[-x^2\frac{1+z}{1-z}\right] \, \mathcal{K}^{(z)}_j(x)
\right\}\bigg{\vert}_{z=0}
\right\},
\end{equation}\normalsize
where
\footnotesize\begin{eqnarray}
\label{finalform2B}
\mathcal{K}^{(z)}_j(x) 
&=&\frac{\sqrt{\pi}}{2} \sum_{\ell = 0}^{\upsilon-1+2j}
\frac{x^{-(2\ell+1)}(\upsilon-1+2j)!}{(\upsilon-1+2j-\ell)!\,\ell!}
\left(\frac{d~}{d\mu}\right)^{\ell}\left\{
\frac{1}{\sqrt{\mu}}
\exp\left(-\frac{k^{2}}{\mu}\right)\, \Re\left[\mbox{Erf}\left(\sqrt{\mu}x + i \frac{k}{\sqrt{\mu}}\right)\right]
\right\}\bigg{\vert}_{\mu=\frac{1+z}{1-z}}.\nonumber
\end{eqnarray}\normalsize

\begin{figure}
\vspace{-1 cm}
\includegraphics[scale=0.43]{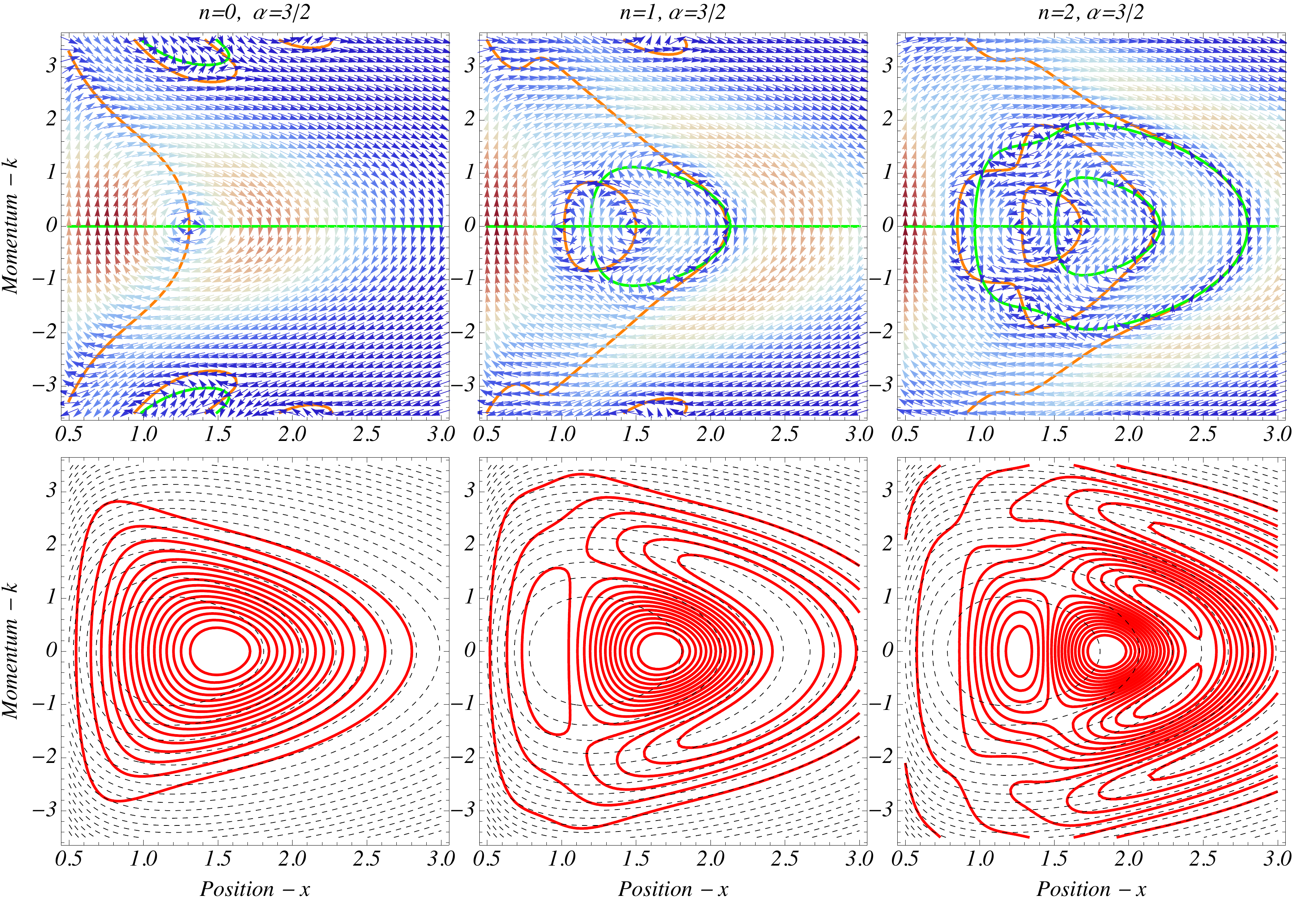}
\includegraphics[scale=0.43]{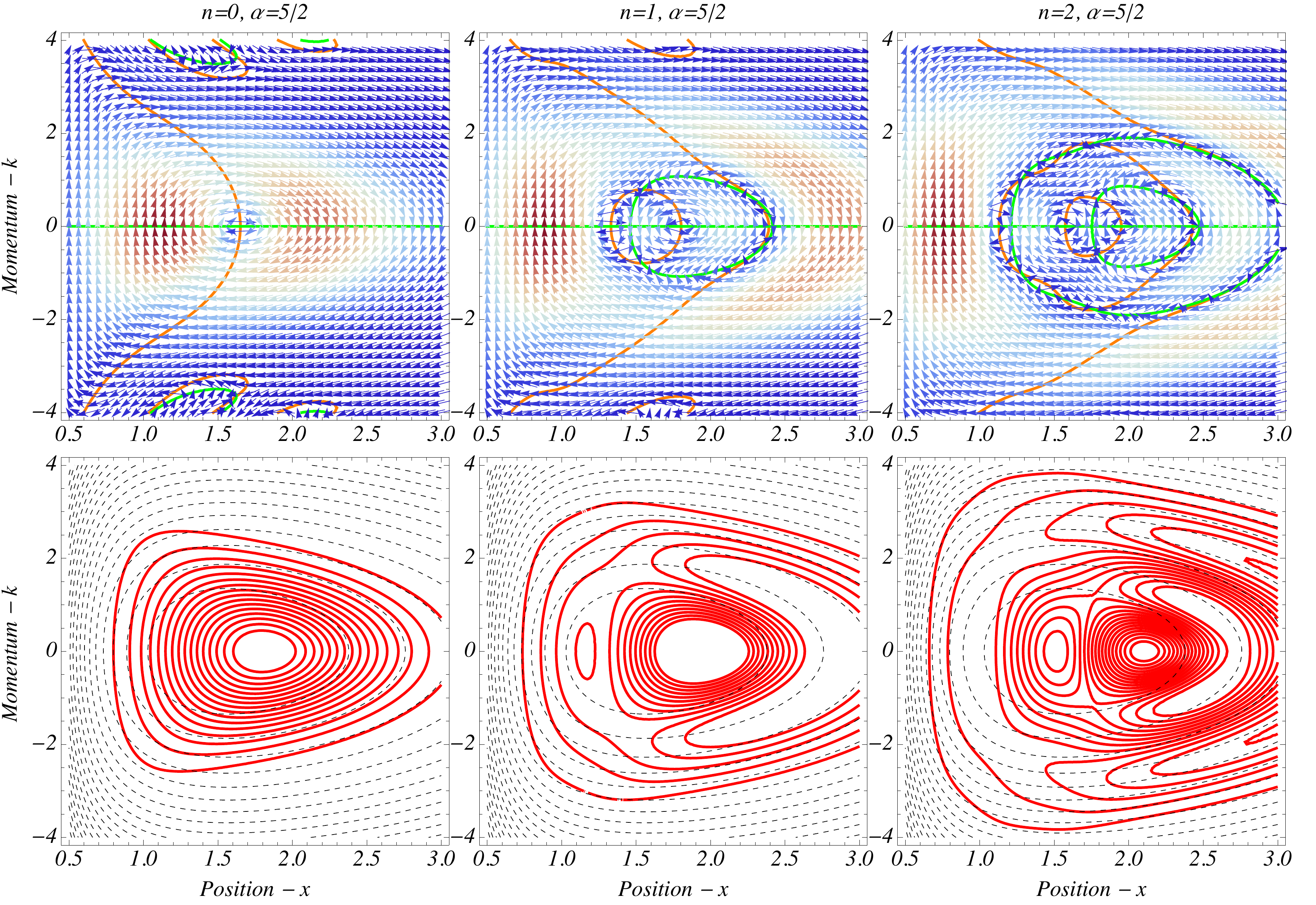}
\renewcommand{\baselinestretch}{.85}
\caption{\footnotesize{
(Color online) (First and third rows) Features of the Wigner flow for $\mathcal{W}_n^{\alpha}(x, \, k)$ in the $x- k$ plane, for quantum numbers $n=0,\,1$ and $2$ (from left to right).
Green contour lines are for $\mathcal{J}_x^{n(\alpha)}(x,\,k) = 0$ and orange contour lines are for $\mathcal{J}_k^{n(\alpha)}(x,\,k) = 0$.
The contour lines are bounds for the reversal of the Wigner current $x$ and $k$ components. Their intersections are stagnation points.
The background {\em thermometer} color scheme (from minimal (blue) to maximal (red) values) describes the modulus of the Wigner current profiles, $\vert\mbox{\boldmath$\mathcal{J}$}^{n(\alpha)}_{k}(x,\,k)\vert$, with the domains of quantum fluctuations bounded by green and orange lines.
(Second and forth rows) Corresponding Wigner function profile (solid red contour lines) and the classical background trajectories (dashed black lines).
The sets of plots are for $\alpha = 3/2$ (first two rows) and $\alpha = 5/2$ (last two rows).}}
\label{Figura001}
\end{figure}

Through the above results, the quantum fluctuations can be identified by the Wigner flow stagnation points at which $\mathcal{J}^{n(\alpha)}_{x}(x,\,k)=\mathcal{J}^{n(\alpha)}_{k}(x,\,k)=0$, as they are depicted in Fig.~(\ref{Figura001}), identified by orange-green crossing lines.
The quantum portrait can be compared with the classical one\footnote{Given by a collection of black lines describing the normalized field equation driven by the vectorial current $$\mbox{\boldmath$\mathcal{J}$}^{n(\alpha)} = (\mathcal{J}^{n(\alpha)}_{x},\mathcal{J}^{n(\alpha)}_{k}) \propto \left(k,\,- x - \frac{1-4\alpha^2}{4x^3}\right),$$ which of course does not depend on the quantum number $n$.} for which the quantum features are completely suppressed by truncating the series expansion, Eq.~(\ref{qua503}), at $\nu=0$.
Of course, the quantum fluctuations have increasing relevant amplitudes for increasing values of $n$.
Otherwise, the increasing value of the parameter $\alpha$ suppresses the quantum effects and approaches the system to the classical profile. It can be evinced by the behavior of the external (red) fringes of the corresponding Wigner functions which, in this case, correspond to the transition of quantum into classical trajectories.
The contributions from stagnation points are identified by (anti)clockwise vortices with winding number equals to $(-)+1$, and by separatrix intersections and saddle flows, both with vanishing winding numbers.
As it has been discussed in the literature \cite{Liouvillian,EPL18,JCAP18}, a set of contra-flux fringes with a domain defined by green and orange lines emerges as a compensating effect which contra-balances the evolution of the quantum flux.
Given that the classical profile does not exhibit such topological fluctuations, the quantifiers of non-classicality proposed in Section II can quantitatively account for these effects along a classical domain delimited by some total energy associated to a classical trajectory, $\mathcal{C}$, as one shall verify in the following.

\subsubsection*{Quantum effects for bounce models}

Before proceeding, one could pay attention to a modified bounce model version of the above discussed problem.
In fact, some fundamental questions in quantum mechanics are reflected onto the discussion of discontinuities and singularities.
In quantum cosmology, for instance, it is related to the formulation and circumvention of the initial singularity problem \cite{Hartle83,Linde84,Rubakov84,Vilen84}.
Discontinuities on the derivative of quantum potentials also affect the exact resolution of wave packet scattering and quantum tunneling subtleties at the standard quantum mechanics.
Through the analytic continuation of the coordinate $x$ from $(0,\infty)$ to $(-\infty,\infty)$, the presence of an infinity potential barrier at $x=0$ constrains the mirror solution as to exhibit an identical behavior of the above resolved problem.
The attenuation of such an infiniteness of the potential barrier, at $x = 0$, means that the wave functions from left to right (and vice-versa) should be probabilistically connected.
A realistic approach for describing such a bounce model scenario can be implemented on the results from Eqs.~\eqref{DimW6} and \eqref{finalform2B} by suppressing the {\em error}-functions from the final results by setting $\mbox{Erf}[\dots]\to 1$. This is equivalent to the elimination of the step-functions, $\Theta(x)$, from the corresponding preliminary integrations.

\begin{figure}
\includegraphics[scale=0.43]{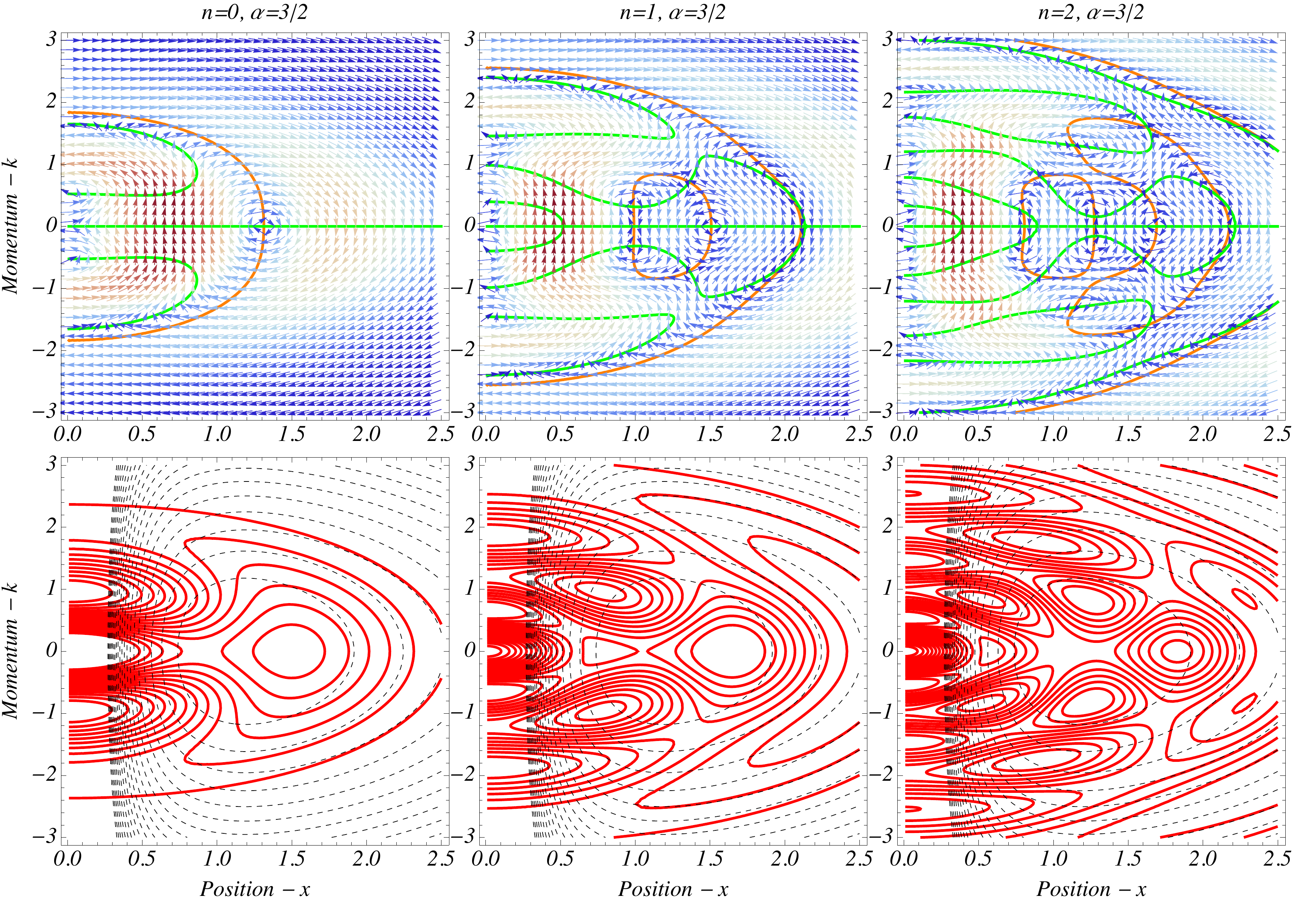}
\includegraphics[scale=0.43]{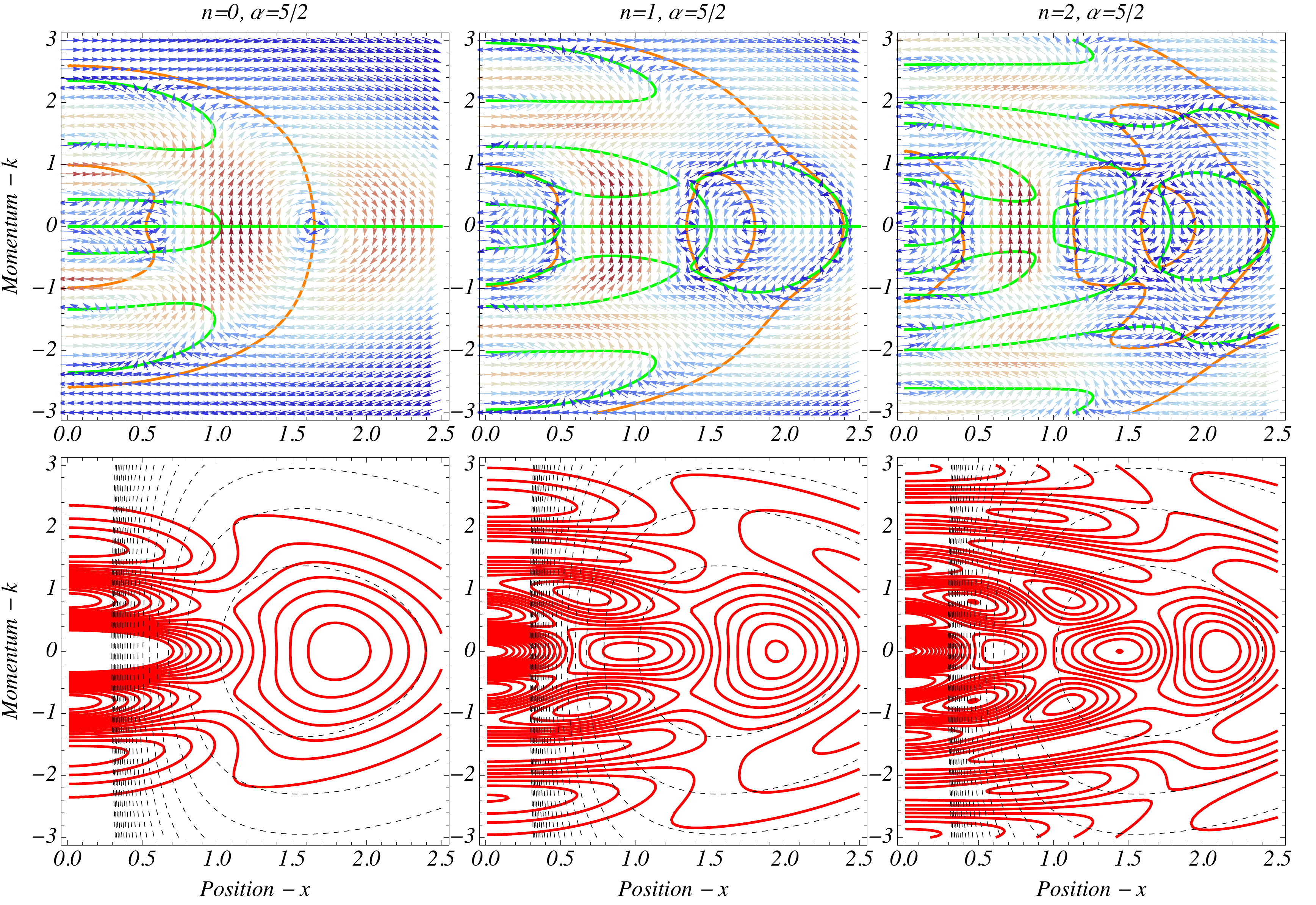}
\renewcommand{\baselinestretch}{.85}
\caption{\footnotesize{
(Color online) Right section ($x > 0$) of the quantum bounce model related to $\mathcal{W}_n^{\alpha}(x, \, k)$, in correspondence with Fig.~\ref{Figura001}, for quantum numbers $n=0,\,1$ and $2$ (from left to right).
Again, the contour lines are bounds for the reversal of the Wigner current $x$ and $k$ components, the color scheme follows the same one from Fig.~\ref{Figura001}, and the plots are for $\alpha = 3/2$ (first two rows) and $\alpha = 5/2$ (last two rows).}}
\label{Figura002}
\end{figure}

In the above context, the bounce model introduces largely suppressed quantum tunneling fluctuations, whose influence can be computed from such a redefined Wigner function, as it can be seen from Fig.~\ref{Figura002}.
The quantum tunneling from left to right (and vice-versa) results into quantum fluctuations near to $|x|
\sim 0$. 

Of course, the inclusion of an artificial (finite) potential barrier at $x=0$ affects the quantumness of the problem due to the stagnation points that emerge at $x=0$. Although the above qualitative prescription can be provided, such a bounced quantum configuration does not fit the (classical) boundary conditions for the application of the flux equations, since a typical classical trajectory cannot be obtained for the approximated bounce model.

\section{Classical {\em versus} quantum portraits}

For the dimensionless Hamiltonian,
\begin{equation}
\mathcal{H}(x,\,k) = \frac{1}{2}\left(k^2+ x^{2}+ \frac{4 \alpha^2 -1}{4 x^{2}} - 2 \alpha\right),
\label{qua40}
\end{equation}
the classical trajectories are resumed by $\mathcal{H}_{\mathcal{C}}= \varepsilon$, where $\varepsilon$ is the dimensionless classical energy. 

The evaluation of the Poisson brackets thus yields
\begin{eqnarray}
\dot{k}_{_{\mathcal{C}}} &=& \,\{k_{_{\mathcal{C}}},\,H\}_{\mbox{\tiny PB}} = - \left(x_{_{\mathcal{C}}}^{2}+ \frac{4 \alpha^2 -1}{4 x_{_{\mathcal{C}}}^{2}}\right),\\
 \dot{x}_{_{\mathcal{C}}} &=& \,\{x_{_{\mathcal{C}}},\,H\}_{\mbox{\tiny PB}} = k_{_{\mathcal{C}}},
\label{qua42}
\end{eqnarray}
where ``{\em dots}'' denote time derivatives. The resolution of the corresponding equations of motion for $x_{_{\mathcal{C}}} \in (0,\infty)$ results into
\begin{eqnarray}
x_{_{\mathcal{C}}}\bb{\tau_{}} &=& \sqrt{\alpha +\varepsilon +  \sqrt{\varepsilon^2 + 2\alpha\varepsilon +1/4}\cos(\tau +\vartheta)},\\
k_{_{\mathcal{C}}}\bb{\tau_{}} &=& \frac{\sqrt{\varepsilon^2 + 2\alpha\varepsilon +1/4}\sin(\tau +\vartheta)}{\sqrt{\alpha +\varepsilon +  \sqrt{\varepsilon^2 + 2\alpha\varepsilon +1/4}\cos(\tau +\vartheta)}},
\label{qua44}
\end{eqnarray}
where $\vartheta$ depends on the initial conditions. The coordinates $x_{_{\mathcal{C}}}\bb{\tau_{}}$ and $k_{_{\mathcal{C}}}\bb{\tau_{}}$ define the classical trajectory, $\mathcal{C}$, that drive the
path integrals for the Wigner information flux quantifiers.

\begin{figure}
\includegraphics[scale=0.45]{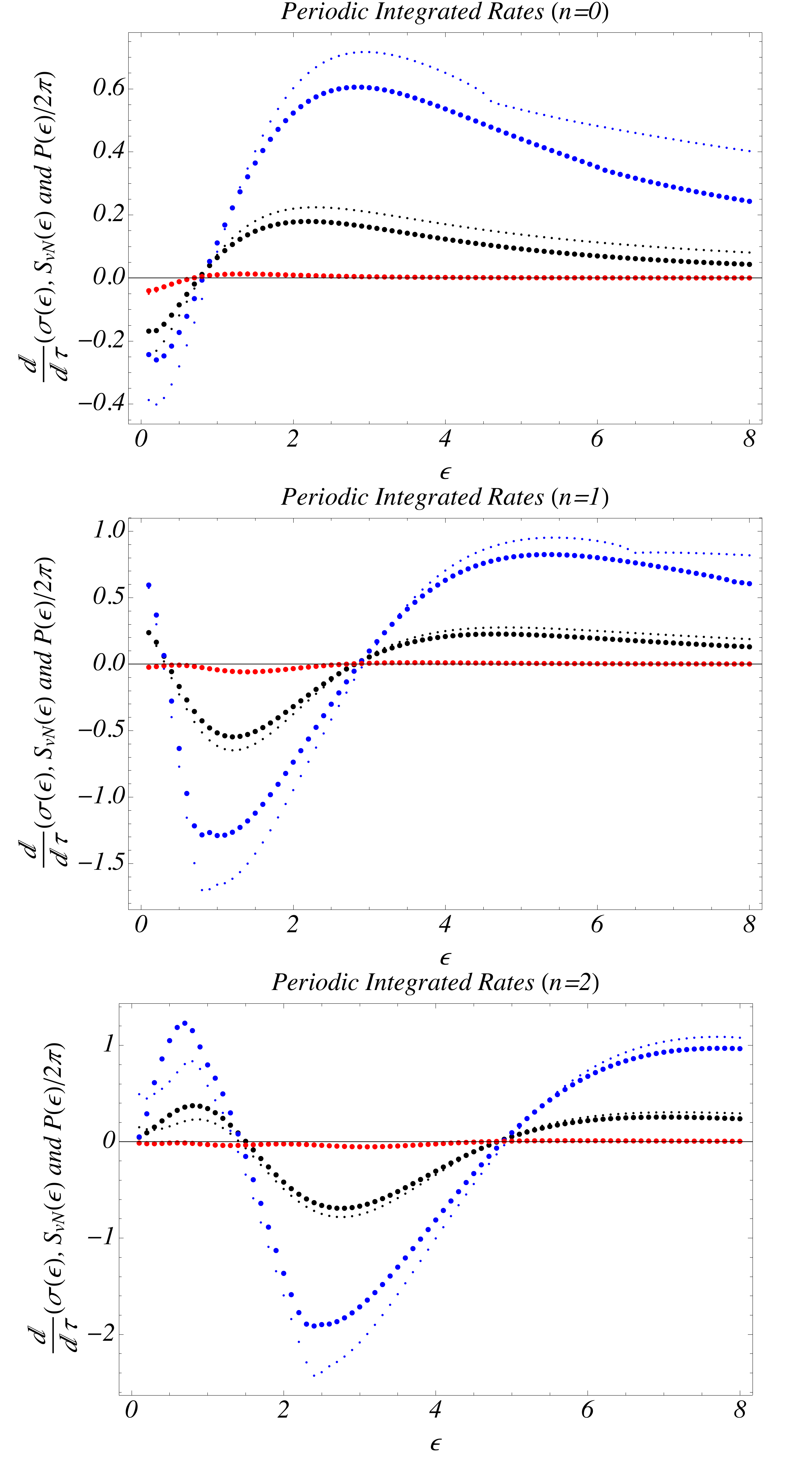}
\renewcommand{\baselinestretch}{.85}
\caption{(Color online) Quantifiers of decoherence (black), entropy flux (blue) and purity flux (red) for the periodic anharmonic system driven the Hamiltonian Eq.~\eqref{qua14} as function of the total energy parameter $\varepsilon$ for quantum states described by $n = 0$ (first plot),  $n = 1$ (second plot) and $n =2$ (third plot), a for $\alpha = 3/2$ (small dots) and $\alpha = 5/2$ (large dots).
The results correspond to the rates of local transference of information throughout the boundary surface, $\mathcal{C}$, respectively expressed by $\frac{D~}{D\tau}\sigma_{(\mathcal{C})}
{\vert}_{\tau = T}$ (black), $\frac{D~}{D\tau}{S}_{vN(\mathcal{C})}
{\vert}_{\tau = T}$ (blue), and $\frac{D~}{D\tau}\mathcal{P}_{(\mathcal{C})}
{\vert}_{\tau = T}$ (red), along a period of motion, $T = 2\pi$.}
\label{Figura003}
\end{figure}

Given that 
\begin{eqnarray}
\Delta\mathcal{J}^{n(\alpha)}_{k}(x,\,k) = \mathcal{J}^{n(\alpha)}_{k}(x,\,k) -\mathcal{J}^{n(\alpha)}_{k(Cl)}(x,\,k)
&=& - \frac{1-4\alpha^2}{4} \left( x \mathcal{Y}^{\alpha}_n(x,\,k) - x^{-3}\mathcal{W}^{\alpha}_n(x,\,k)\right),\,\,\,\,
\label{qua505G}
\end{eqnarray}
for the periodic motion along $\mathcal{C}$ defined by a fixed energy, $\varepsilon$, (cf. Eqs.~(\ref{qua42})-(\ref{qua44})), the local features of non-classicality can be computed in terms of integrated periodic probability fluxes enclosed by the two-dimensional boundary surface, $\mathcal{C}$, obtained from Eqs.~\eqref{quaz51EE}, \eqref{quaz62CC}, and \eqref{quaz64DD}, respectively written as
\begin{eqnarray}
\frac{D~}{D\tau_{}}\sigma_{(\mathcal{C})}
\bigg{\vert}_{{\tau_{}} =2\pi }
&=& -
\int_{0}^{2\pi} d\tau_{}\, \Delta\mathcal{J}^{n(\alpha)}_{k}(x_{_{\mathcal{C}}}\bb{\tau_{}},\,k_{_{\mathcal{C}}}\bb{\tau_{}})\,\,k_{_{\mathcal{C}}}\bb{\tau_{}},\\
\frac{D~}{D\tau_{}}{S}_{(\mathcal{C})}
\bigg{\vert}_{{\tau_{}} = 2\pi}
&=& 
\int_{0}^{2\pi} d\tau_{}\, \ln\left[\mathcal{W}(x_{_{\mathcal{C}}}\bb{\tau_{}},\,k_{_{\mathcal{C}}}\bb{\tau_{}})\right]\,\Delta\mathcal{J}^{n(\alpha)}_{k}(x_{_{\mathcal{C}}}\bb{\tau_{}},\,k_{_{\mathcal{C}}}\bb{\tau_{}})\,\,k_{_{\mathcal{C}}}\bb{\tau_{}},\\
\frac{1}{2\pi}\frac{D~}{D\tau_{}}\mathcal{P}_{(\mathcal{C})}
\bigg{\vert}_{{\tau_{}} = 2\pi}
&=& -
\int_{0}^{2\pi} d\tau_{}\, \mathcal{W}(x_{_{\mathcal{C}}}\bb{\tau_{}},\,k_{_{\mathcal{C}}}\bb{\tau_{}})\,\Delta\mathcal{J}^{n(\alpha)}_{k}(x_{_{\mathcal{C}}}\bb{\tau_{}},\,k_{_{\mathcal{C}}}\bb{\tau_{}})\,\,k_{_{\mathcal{C}}}\bb{\tau_{}},
\label{quaz62CCC}
\end{eqnarray}
for which the results are depicted in Fig.~\ref{Figura003}.
Given that increasing values of $\alpha$ approach Wigner to classical profiles, as one can notice from Fig.~\ref{Figura001}, it should be natural to expect such a correspondence with the results from Fig.~\ref{Figura003}, where the overall averaged amplitude of the non-classicality fluxes are suppressed for larger values of $\alpha$.

The above triplet describing the fluxes of information are totally consistent with each other. Their associated integrated quantifiers all depict the equivalent rates of quantum discrepancies from classical regimes parameterized by $\mathcal{C}$.
One also notices that the associated energy parameter, $\varepsilon$, in correspondence with the quantum energies, $\varepsilon_n = 2n+1$, reproduces a kind of Bohr-Sommerfeld quantization scheme identified for the fluxes of information.
For increasing values of $\varepsilon$ (and $n$), such that the quantum distortions are homogenized according to the phase-space volume considered, the quantum regime is identified by the mutual crossing (at zero nodes) of all quantifiers at $\varepsilon = \varepsilon_n = 2n+1$ largest value.
As preliminarily reported for anharmonic P\"oschl-Teller quantum potentials \cite{EPL18}, phase-space classical trajectories only accommodate (without yielding quantum distortions) the corresponding $n$-quantized version of the quantum system.
The nodes indicate that sum of winding numbers related to the quantum stagnation points enclosed by $\mathcal{C}$ average out to zero. Of course, the deviations from quantizing trajectories for arbitrary values of $\varepsilon$ have an evident correspondence with the flux of quantum information through the classical boundary, $\mathcal{C}$, properly quantified in Fig.~\ref{Figura003}.

\section{Conclusions}

A fluid analog of the phase-space information flux related to purity and von Neumann entropy, once driven by Wigner functions and Wigner currents, has been associated to the already known quantum decoherence and non-Liouvillian aspects of quantum systems \cite{EPL18,Steuernagel3,Ferraro11,Donoso12,Domcke}.
In this work, this framework has been extended to the investigation of the quantum system driven by the harmonic oscillator potential modified by an inverse square (one dimension Coulomb-like) contribution, for which exact expressions for Wigner functions and Wigner currents have been obtained.
In this context, quantumness and classicality given in terms phase-space quantum decoherence, purity and von Neumann entropy fluxes have been again investigated in order to extend a preliminary discussion recently introduced for hyperbolic quantum wells \cite{EPL18} and quantum cosmological scenarios \cite{JCAP18}.
Also relevantly, assuming that some mathematical manipulability of the Weyl transformed associated to (arbitrary) quantum states is identified, and that the corresponding quantum potentials support a periodic motion which defines an enclosing phase-space classical trajectory, our results are consistent with the assertion that this Wigner flow framework can be universally applied to any one-dimensional periodic physical system in order to evaluate how quantum regimes are far from classical ones.

{\em Acknowledgments} -- This work was supported by the Brazilian agencies FAPESP (grant 2018/03960-9) and CNPq (grant 300831/2016-1).

\end{document}